\documentclass[
aps,
prab,
reprint,
twocolumn,
superscriptaddress]
{revtex4-2}

\usepackage{amssymb}
\usepackage{amsthm}
\usepackage{fancyhdr}
\usepackage{graphicx}
\usepackage[colorlinks, linkcolor=green!40!black, citecolor=blue!80!black]{hyperref}
\usepackage[table]{xcolor}
\usepackage{multirow}
\usepackage{rotating}
\usepackage{float}
\usepackage{array}
\usepackage{supertabular}
\usepackage{amsmath}
\usepackage{dcolumn}
\usepackage{tipa}
\usepackage{bm}
\usepackage{booktabs}
\usepackage[capitalise]{cleveref}
\usepackage{optidef}
\usepackage{tabularx}
\usepackage{orcidlink}
\usepackage[T1]{fontenc}
\usepackage{amsmath}
\usepackage{natbib}
\usepackage{appendix}
\usepackage{float}
\usepackage{soul}
\usepackage{xcolor}

\newcommand{\aap}{Astron. Astrophys.}
\newcommand{\mnras}{Mon. Not. Roy. Astron. Soc.}

\begin{document}

\title{A population study on the effect of metallicity on ZAMS to the merger}

\author{Sourav Roy Chowdhury \orcidlink{0000-0003-2802-4138}}
\email{roic@sfedu.ru}
\affiliation{Research Institute of Physics, Southern Federal University, 344090 Rostov on Don, Russia.}
\affiliation{Department of Physics, Vidyasagar College, 39, Sankar Ghosh Lane, Kolkata, India.}
\author{Deeptendu Santra \orcidlink{0000-0001-5433-3351}}
\email{research@dsantra.com}
\affiliation{Institute of Engineering and Management, Sector-5, Kolkata, India.}

\date{\today} 

\begin{abstract}
The formation channels of compact object binaries are crucial for interpreting gravitational wave observations and enhancing early multi-messenger alerts.
Despite the key role of stellar metallicity in progenitor evolution, many models assume a uniform value for all stars in a cluster.
In this study, we investigate the impact of a heterogeneous stellar metallicity distribution on the formation of compact object binaries and their resulting gravitational wave signatures.
We extend the COSMIC binary population synthesis code to incorporate a fiducial mass-redshift-metallicity relation for individual Zero-Age Main Sequence stars of a stellar cluster.
Focusing on low-metallicity environments at redshift $z \sim 4$, we analyse the gravitational-wave signals from binary black holes and black hole–neutron star systems in the sub-hertz regime.
The resulting binary black holes have total masses from 8-86 \rm{M}$_\odot$, while black hole–neutron star systems range from 6-31 \rm{M}$_\odot$.
We assess the detectability of the characteristic strains against the sensitivity curves of planned sub-hertz observatories.
\end{abstract}

\maketitle

\section{Introduction} \label{sec:intro}


The detection of gravitational waves (GWs) has revolutionised our understanding of the demographics and physical properties of compact object (CO) mergers. As CO binaries inspiral and merge, they emit gravitational radiation observable by GW detectors. Advanced LIGO, Virgo, and KAGRA (LVK) Collaboration has observed numerous GW events \citep{Abbott2,Abbott3,Abbott4}, beginning with the landmark detection of GW150914 \citep{Abbott1}. Since then, nearly 100 events\footnote{\url{https://www.gw-openscience.org/eventapi/html/GWTC/}} have been recorded across multiple observing runs. However, GW signals from other astrophysical sources, such as white dwarf binaries \citep{wd}, magnetars \citep{magnetars1, magnetars2}, core-collapse supernovae \citep{ccsne2,ccsne1}, primordial BH \citep{pbh1,pbh2,pbh3}--remain undetected and are critical targets for next-generation observatories.


A range of evolutionary channels has been proposed to explain the diversity of GW events detected by the LVK Collaboration. Observational studies suggest that massive zero-age main sequence (ZAMS) stars, susceptible to electron-positron pair-instability \citep{pair_ins1,pair_ins2}, may either be completely disrupted or undergo pulsational mass loss, eventually forming COs. Resulting black holes (BHs) typically have final masses up to 50~\rm{M}$_\odot$, while neutron stars (NSs) masses up to 2.5~\rm{M}$_\odot$ \citep{Fryer,marchant_m}. These outcomes highlight binary evolution as a key pathway for CO formation \citep{sana1, Moe_17}. Isolated binary evolution is a leading channel \citep{mt2,mt1}, wherein massive binaries undergo mass transfer and common-envelope phases, forming CO binaries that coalesce within Hubble time \citep{SN3,SN1,SN4,SN2}. Alternative scenarios include dynamical formation in dense environments via multi-body interaction \citep{dyn_evo3}; chemically homogeneous evolution, where rapid rotation leads to efficient mixing and prevents stellar expansion \citep{che_evo1,che_evo3}.


In all these channels, the metallicity of the progenitor stars plays a decisive role in determining the physical properties of the remnants and the evolutionary pathways of their host clusters. Metallicity, in particular, has been recognised as a fundamental parameter influencing stellar winds, mass loss rates, remnant masses, and the likelihood of forming different types of compact objects \citep{Ziosi,mapelli_m,Boco_2021,marchant_m}. Low-metallicity environments lead to reduced stellar wind mass loss, enabling the formation of more massive BHs compared to metal-rich populations \citep{Mapelli_2020,Lin,vink} and increasing the fraction of BH progenitors in stellar populations \citep{Giacobbo_co,sevn1}. This, in turn, affects the predicted merger rates and mass distributions of CO binaries, which are key observables for GW astronomy\citep{met_sf,met_gw,di_co}. To accurately capture these effects, modern stellar evolution and population synthesis codes explicitly model metallicity-dependent processes \citep{met_code1,met_code2}.


Stellar evolution from ZAMS to CO formation is typically modelled using a combination of simulations and analytical prescriptions, providing a notion for understanding evolutionary pathways. Several codes have been developed to model the evolution of individual stars and binaries systematically. MESA \citep{mesa1,mesa2} is a modular 1D stellar evolution code primarily for single stars, with binary capabilities. BINSTAR \citep{bin2} and EV/STARS/TWIN \citep{bps7,bps6} focus on binary evolution, modelling interactions such as mass transfer, Roche-lobe overflow, common envelope evolution, chemical composition, wind mass loss and so on. Binary Population Synthesis (BPS) methods integrate single-star evolution with binary interaction prescriptions to simulate stellar populations from the ZAMS to their remnant stages. BPS methods were initially introduced by \cite{bps1} and have since been refined for a variety of astrophysical contexts. Population synthesis codes such as SEVN \citep{sevn1}, COMBINE \citep{com1}, or POSYDON \citep{posy,posy2} generate extensive stellar evolution tables, applying interpolation alongside simplified binary evolution models. Other approaches like StarTrack \citep{StarTrack} COMPAS \citep{comp} utilise analytic fitting formulas from \cite{hurley} and remnant mass prescriptions from \cite{Fryer}. BPASS \citep{bpass1} relies on pre-computed stellar models derived from the STARS/EV code coupled with grid interpolation. The semi-analytical code B-POP \citep{BPOP} specifically aims to predict properties and merger rates of GW sources. The TrES code \citep{tres} models the complex dynamical interactions within three-body systems, while DRAGON-II provides a suite of 19 direct N-body simulations \citep{Dragon1,Dragon2} specifically designed to study hierarchical mergers.


Interpreting GW events from CO mergers requires a comprehensive notion of their formation history, informed by observations, theory, and simulations across diverse evolutionary scenarios. ZAMS stars are expected to form COs through distinct channels, which may later merge as GW-emitting binaries.
\citep{Langer,dyn_evo1,cosmic,comp,Banerjee2021b,MANDEL2021}. Modelling CO formation and evolution using BPS and delay-time distributions has provided key insights into GW event rates. Notably, the \cite{gw1} kick model closely reproduces the observed chirp mass distribution of stellar-origin mergers. \cite{gw3} applies delay-time models to a single stellar population, while \cite{zeist_gw} analyses integrated GW spectra from model clusters with varied compact binary populations. Post-merger BH retention depends on gravitational recoil kicks, which are mitigated in isotropic configurations-enabling hierarchical mergers in dense clusters \citep{Miller2002, Rodriguez2016b}. Such a mechanism is invoked to explain events such as GW190521 \citep{Fragione2021, Tagawa2021}. Looking ahead, future GW observatories [Einstein Telescope, Cosmic Explorer, LISA, LGWA \citep{Reitze2019,Rodriguez2019,Maggiore_2020,lisa,lgwa}] will play a pivotal role in disentangling these channels and probing the redshift evolution of compact binaries.

Despite the recognised importance of metallicity, many BPS models still apply a single, uniform metallicity to an entire stellar cluster. In this work, we explore the consequences of relaxing this constant-metallicity assumption. Using the COSMIC \citep{cosmic_software}, we evolve large populations of binary stars from specified initial conditions, assigning metallicities to each star individually based on an empirical mass–redshift–metallicity relation. Our study focuses on low-metallicity star clusters forming at $z \sim 4$, a regime where sub-solar abundances strongly influence the masses, merger rates, and GW signatures of compact remnants. We simulate the formation and evolution of ZAMS binaries into their first-generation COs-both binary black holes (BBHs) and black hole-neutron star (BH-NS) systems—and compute their characteristic strains in the sub-Hz frequency range. In this frequency regime, the convergence timescale for CO binaries can range from months to millions of years \citep{subhz}, making them important long-lived inspiral targets. For simplicity, we consider that the stellar objects have zero spin at birth. The structure of the paper is as follows: Section \ref{sec2} details the methodology, including our metallicity prescription and GW strain calculation. Section \ref{sec3} presents the simulation results, tracing binary evolution from progenitors to remnant mass distributions and expected GW signals. Section \ref{sec4} provides a summary and discussion of our findings.

\section{Methodology}\label{sec2}

Stellar clusters forming at high redshift provide a crucial laboratory for understanding the role of early-Universe environments in the evolution of COs.
The low-metallicity conditions characteristic of this epoch offer a unique framework for probing the processes that govern CO progenitors. 
Metallicity is a particularly critical parameter, as it directly influences stellar wind mass loss, supernova mechanisms, and binary interaction pathways, which together shape the orbital properties and merger outcomes of these systems.

To realistically model these environments, we introduce a fiducial metallicity model that imposes a distribution of progenitor metallicities consistent with star-forming environments at $z \sim 4$  \citep{z4_1, z4_2, z4_4}. This allows us to generate binary populations that accurately reflect the chemical composition of the early Universe (discussed further in Section~\ref{sec:metallicity}).
To implement this model and evolve the binary systems, we use a customised version of COSMIC v3.6.1, which we have enhanced to incorporate our fiducial model into its initialisation routine.

The section is organized as follows:
\begin{itemize}
    \item Introduce our fiducial model of metallicity.
    \item Define the model parameters for generating the initial conditions for the BPS code and evolve the ZAMS for various metallicity conditions.
    \item Determine the characteristics strain of GWs from 1st Gen COs (The first Gen COs are those formed from ZAMS stars).
\end{itemize}

\subsection{Metallicity}\label{sec:metallicity}

Stellar mass and metallicity are key parameters whose correlation is central to understanding the physical processes regulating CO formation in stellar clusters. Stars on the ZAMS are predominantly composed of hydrogen \citep{Norberg-Lecture4}. The number density ratio of oxygen to hydrogen in the gas is commonly expressed as $\log_{10}$[O/H] + 12, known as the solar abundance \citep{meta_rela2}. The choice of solar abundance directly affects metallicity calibrations. This correlation is particularly significant in GW astrophysics, as binary formation in low-metallicity environments strongly influences the GW source population \citep{Lamberts,Chruslinska}.

Metallicity of a progenitor star at the ZAMS stage critically determines its evolutionary endpoint, primarily through its impact on stellar mass loss \citep{Giacobbo_co,Shepherd:2025djj}. Metal-rich stars experience enhanced radiation-driven winds due to higher opacities, resulting in substantial mass loss that often precludes BH formation and instead yields NSs or, in some cases, no compact remnant \citep{mmr1,mmr2,mmr3}. In contrast, metal-poor stars retain more mass and are more likely to undergo direct collapse into BHs \citep{mmr5,met_gw,meta_rela1}.  As a result, high-metallicity progenitors are more likely to produce NSs with remnant masses below the Tolman–Oppenheimer–Volkoff limit \citep{tov2,tov}, whereas low-metallicity progenitors tend to retain more mass and are thus more favourable sites for BH formation via direct collapse \citep{mmr6}. This metallicity dependence is supported by both theoretical models and observational data, including results from JWST and GW detections, which indicate that binary BBHs predominantly originate in low-metallicity environments \citep{mmr7,met_gw,sevn1,met_poor_1,met_poor_2,mmr9}.

Observational and theoretical studies consistently demonstrate a decline in metallicity with increasing redshift, reflecting the progressive chemical enrichment \citep{ma,mass_met_z}. Spectroscopic surveys and absorption line analyses confirm that early galaxies and Damped Lyman-$\alpha$ systems were significantly more metal-poor than their low-redshift counterparts \citep{damped_Lya}. Moreover, metallicity decreases with increasing redshift, reflecting the trend of cosmic chemical evolution \citep{ma} and metal-poor gas accretion in the early universe \citep{mmr10}.

\begin{figure}[h!]
    \centering
    \includegraphics[scale=0.9]{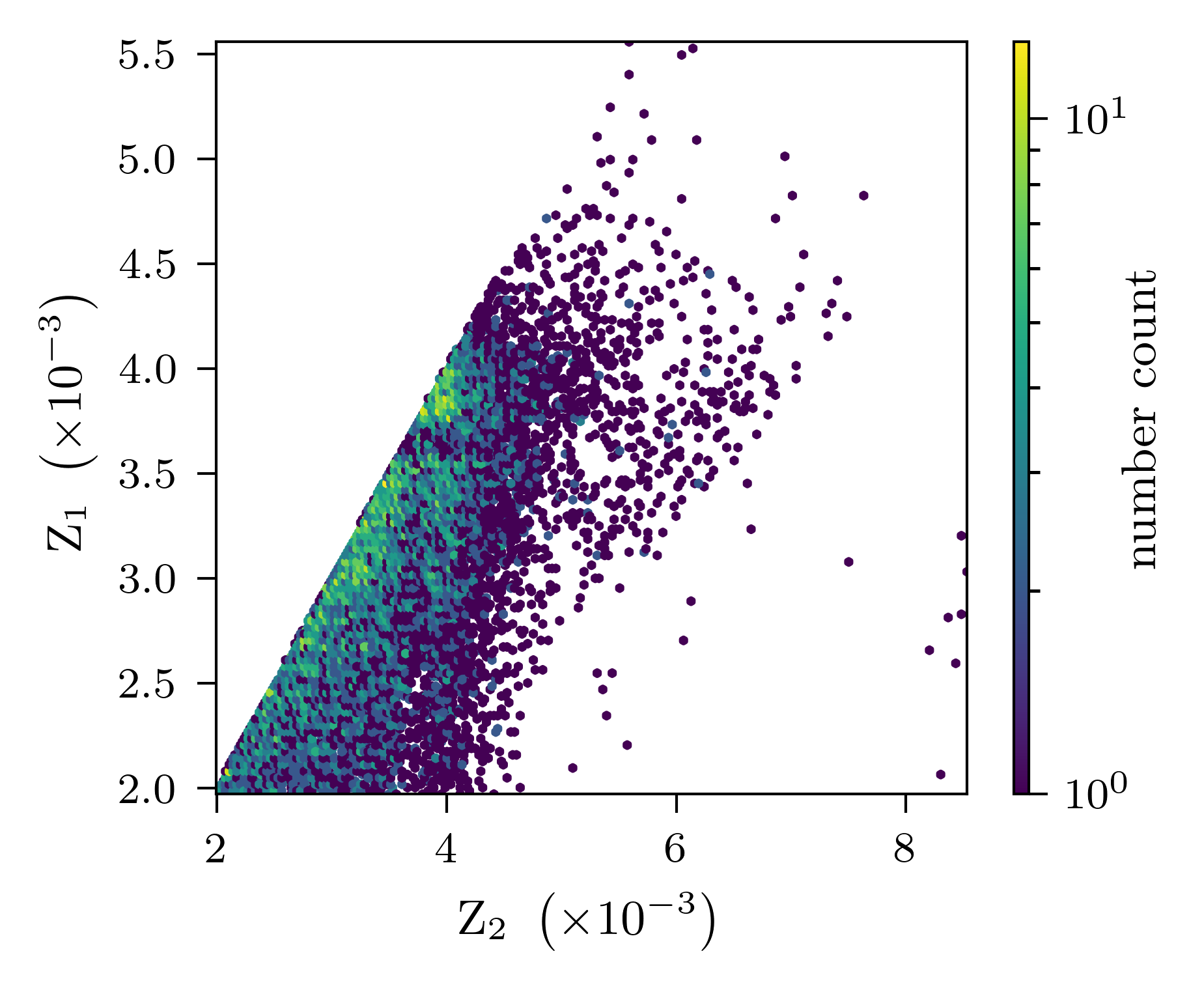}
    \caption{Scatter density plot of the metallicities of the primary (\rm{Z}$_1$) and secondary (\rm{Z}$_2$) stellar progenitors in binary systems. The colour scale represents the number count of progenitor systems.}
    \label{met_var}
\end{figure}

Motivated by these findings, we adopt an empirical relation to describe the dependence of stellar metallicity ($\mathrm{Z}$) on both the redshift of the stellar cluster and the mass of individual progenitor stars:
\begin{equation} \label{e1}
\log_{10}\left(\frac{\mathrm{Z}}{\mathrm{Z}_\odot}\right) = p\log_{10}\left(\frac{\mathrm{M}_\odot}{\mathrm{M}}\right) + q\exp(-rz),
\end{equation}
where $p, q \textrm{~and~}r$ are the correlation coefficient (set to $p=0.5$ for simplicity and $q = 0.67 \textrm{~and~}r = 0.5$ following \cite{ma}), $m$ is the progenitor mass, and $z$ is the cluster redshift. The first term captures the mass dependence, while the second term models the observed decrease in metallicity with redshift. Since stellar clusters typically lie at a fixed redshift, the second term acts as a constant for a given system.

To implement the low metallicity condition in the ZAMS stars, we set an upper bound on metallicity as $\mathrm{Z}_{max} \leq 0.01$ \citep{bh1,met_eve}. For various combinations of mass and redshift, metallicity can often surpass the considered threshold  $\mathrm{Z}_{max}$. At the considered redshift, any stellar mass exceeding approximately 7.97 M$_\odot$  will maintain a metallicity below $\mathrm{Z}_{max}$.
And it perfectly complements our previous inference about redshift and mass. Therefore, we adopt the M-Z relation in Eq. (\ref{e1}) as a simple and effective model for our simulated binaries. Figure \ref{met_var} represents the scatter densities of the metallicity for the primary and secondary progenitors in a stellar cluster.

As discussed earlier, the original COSMIC takes in a single metallicity value for every binary in the simulation.
We modify the code to allow users to control how metallicity is initialised in the ZAMS.
With the modified code, users can pass in \texttt{metallicity} = \texttt{"individual"} instead of a \texttt{float} value, to use Eq. (\ref{e1}) for computing the metallicities.

The \verb|metallicity| argument can take in more types of values, which is further discussed in APPENDIX \ref{appen}. The complete codebase, along with all the simulation parameters used in this study, is publicly available \footnote{\url{https://github.com/Dsantra92/COSMIC-GW}}.

\subsection{Stellar Population Synthesis}\label{sec:bps}

We initialise ZAMS binaries using COSMIC's \texttt{independent} sampler, wherein stellar parameters are assumed to be uncorrelated.
The primary stellar mass is drawn from the \citet{Salpeter} IMF, while the secondary mass is determined using a uniform mass-ratio distribution \citep{mazeh1992mass, goldman1994orbital}.
To restrict the study to CO (BBH and BH-NS), we impose a minimum mass of $8~\mathrm{M}_\odot$ for the primary and secondary stars, consistent with the threshold for core-collapse supernovae \citep{Hopkins}.
Orbital separations and eccentricities are sampled following \citet{sana1} prescription.
The simulated star formation history starts from 10~\rm{Gyr} \citep{bh1}.
We sample a total of $10^6$ binary systems, sufficient to ensure statistical convergence.
Finally, we use Eq. (\ref{e1}) to assign metallicity values for each star in the system.

To follow the evolution of these ZAMS binaries with enriched metallicity values, we generalised the code so that the metallicity, \(Z\), is stored as a two-element vector rather than a single scalar.
Each component of the binary therefore retains its own chemical composition, which is propagated through all subsequent calculations—the tabulated ignition constants, the Eddington-limited accretion rate, and the metallicity-dependent stellar-wind prescriptions etc..
As the donor and accretor receive distinct ignition parameters, the integrator takes more accurate time-steps and yields a higher-fidelity description of the binary’s evolution.

For the binary stellar evolution, Wind prescriptions follow \citep{vink2005_lbv} with a Reimers mass-loss coefficient of 0.5 \citep{kudritzki1978absolute}. 
Common envelope evolution is governed by an efficiency parameter of 1.0, a variable lambda prescription where the ionisation energy contributes to envelope ejection, and a model for unstable mass transfer based on \citep{hjellming1987thresholds} for giant branch stars. A binary fraction of 0.94 is assumed. Natal kicks for supernovae are implemented using the prescription from \citep{giacobbo2020revising}, with a Maxwellian dispersion of 265 km/s and the inclusion of kicks for BH formation with fallback-modulated kicks.
The remnant mass prescription is based on the delayed model from \citep{Fryer}, setting a maximum neutron star mass of 2.5~\rm{M}$_\odot$ and allowing for pulsational pair-instability supernovae based on \citep{marchant2019pulsational} fits. Mass transfer is Eddington-limited with specific angular momentum loss from the primary. Tidal circularisation is activated using the StarTrack \citep{belczynski2008compact} method. We use solar metallicity, \rm{Z}$_\odot$ = 0.0196 following \cite{solar_met2} and \cite{solar_met1}.

\subsection{Gravitational wave}

The characteristic strain, which includes the effects of an inspiralling signal of $ n^{th} $ harmonic for eccentric and chirping source, can be defined as \citep{s44},
\begin{equation}
h_{c,n}^2 = (\pi D_L)^{-2} \Bigg(\frac{2 G}{c^3} \frac{\dot{E_n}}{\dot{f_n}}\Bigg),\label{e2}
\end{equation}
where $ D_L $ is the luminosity distance of the binary system from the detector. $ \dot{E_n} $ is the power radiated, and $ \dot{f_n} $ is the rate at which the source frame GWs frequency $ (f_n) $ is changing of the $ n^{th} $ harmonic. 

It is often convenient to express the source frame GW frequency in terms of orbital frequency ($ f_o $), which is related by
\begin{equation}
f_n = nf_o.\label{e3}
\end{equation}
where, from Kepler's third law
\begin{equation}
f_o = \Biggl\{\frac{G(\mathrm{M}_1+\mathrm{M}_2)}{(2 \pi a^{3/2})^2}\Biggr\}^{1/2}\label{e4}
\end{equation}
Here, the primary and secondary masses of a CO binary are $ \mathrm{M}_1 $ and $ \mathrm{M}_2 $, respectively; and $ a $ is the semi-major axis of the system.

Following \cite{s43}, the time derivative of the energy radiated in GWs is given by
\begin{equation}
\dot{E_n}(\mathcal{M}_c,f_o,e) = \frac{32}{5} \frac{G^{7/3}}{c^5}(2 \pi \mathcal{M}_c f_o)^{10/3}g(n,e).\label{e5}
\end{equation}
where $\mathcal{M}_c$ is the chirp mass and
\begin{eqnarray}
g(n,e) && = \frac{n^4}{32} \Big[\bigl\{J_{n-2}(ne)-J_{n+2}(ne)-2e\big\{J_{n-1}(ne)\nonumber \\ && -J_{n+1}(ne)\big\}+\frac{2}{n}J_{n}(ne)\bigr\}^2 +(1-e^2)\bigl\{J_{n-2}(ne)\nonumber \\  && \hspace{1cm}  +J_{n-2}(ne)-2J_{n}(ne)\bigr\}^2+\frac{4}{3n^2}[J_{n}(ne)]^2 \Big]\nonumber
\end{eqnarray}
Here, $ J_{n}(ne) $ denotes the first kind of Bessel function.

The rate of variation of the $ n^{th} $ harmonic frequency due to GWs inspiral can be obtained from Eq. (\ref{e2}) and Eq. (\ref{e3}), following the evolution of the semi-major axis with time defined by  \cite{s42}, as
\begin{eqnarray}
\dot{f_n}(\mathcal{M}_c,f_o,e) = \frac{48 n}{5  \pi c^5} (G \mathcal{M}_c)^{5/3} (2 \pi f_o)^{11/3}F(e). \label{e6}
\end{eqnarray} 

GW emission from an eccentric binary, relative to a circular binary with comparable parameters, is quantified by the sum of $g(n, e)$ over all harmonics \citep{s43} as
\begin{equation}
F(e)= \frac{1+\frac{73}{24}e^2+\frac{37}{96}e^4}{(1-e^2)^{7/2}} = \sum_{n=1}^{\infty} g(n,e) \nonumber
\end{equation} 

Now, the definitions of $ \dot{E_n} $ and $ \dot{f_n} $ can be used to plug into Eq. \ref{e2} to derive an expression of the characteristic strain in the detector frame: 
\begin{equation}
h_{c,n}^2 = \frac{2^{5/3}}{3 \pi^{4/3} c^3 D_L^2} \frac{(G \mathcal{M}_c)^{5/3}}{ f_o^{1/3} n^{2/3} (1+z)^2} \frac{g(n,e_n)}{F(e_n)}.\label{e7}
\end{equation}

The maximum number of harmonics of an eccentric binary system is associated with the eccentricity of the system \citep{Wen_2003,nmax} as follows,
\begin{equation}
n_{max}= m\times n_{peak} = 2 \frac{(1+e)^{1.1954}}{(1-e^2)^{3/2}}. \label{e8}
\end{equation}

We have considered here $m=1$. Each of these harmonics contributes to the signal.
The orbital frequency determines the lowest frequency and the highest frequency is determined by the maximum frequency reached at the end of the observation, $f_{o,max} = f_{o} (t = T_{obs})$. Following \cite{Mandel} the GW frequency of these binaries at a time $T_{obs}$ before the merger
\begin{eqnarray}
    f_{o}&& =\frac{n}{32\pi (G \mathcal{M}_c)^{5/8}}\Big(\frac{5 c^5}{T_{obs}}\Big)^{3/8} \nonumber \\&&[(1+0.27e_0^{10}+0.33e_0^{20}+0.2e_0^{1000})(1-e_0^2)^{7/2}]^{3/8}.~~~~
\end{eqnarray}

\section{Result}
\label{sec3}
We now examine how our metallicity-aware population synthesis impacts the demographics of compact object binaries and their observable GW signals. Within the scope of state-of-the-art cosmological simulations, we can take into account the M-Z relation of individual binaries. The simulation reveals that the evolution of CO is strongly governed by the M–Z properties of its progenitors. In this section, we present the outcomes of the simulation. 
\subsection{Initial binaries}\label{result_init_bin}
\begin{figure}[htb!]
  \centering
 \includegraphics{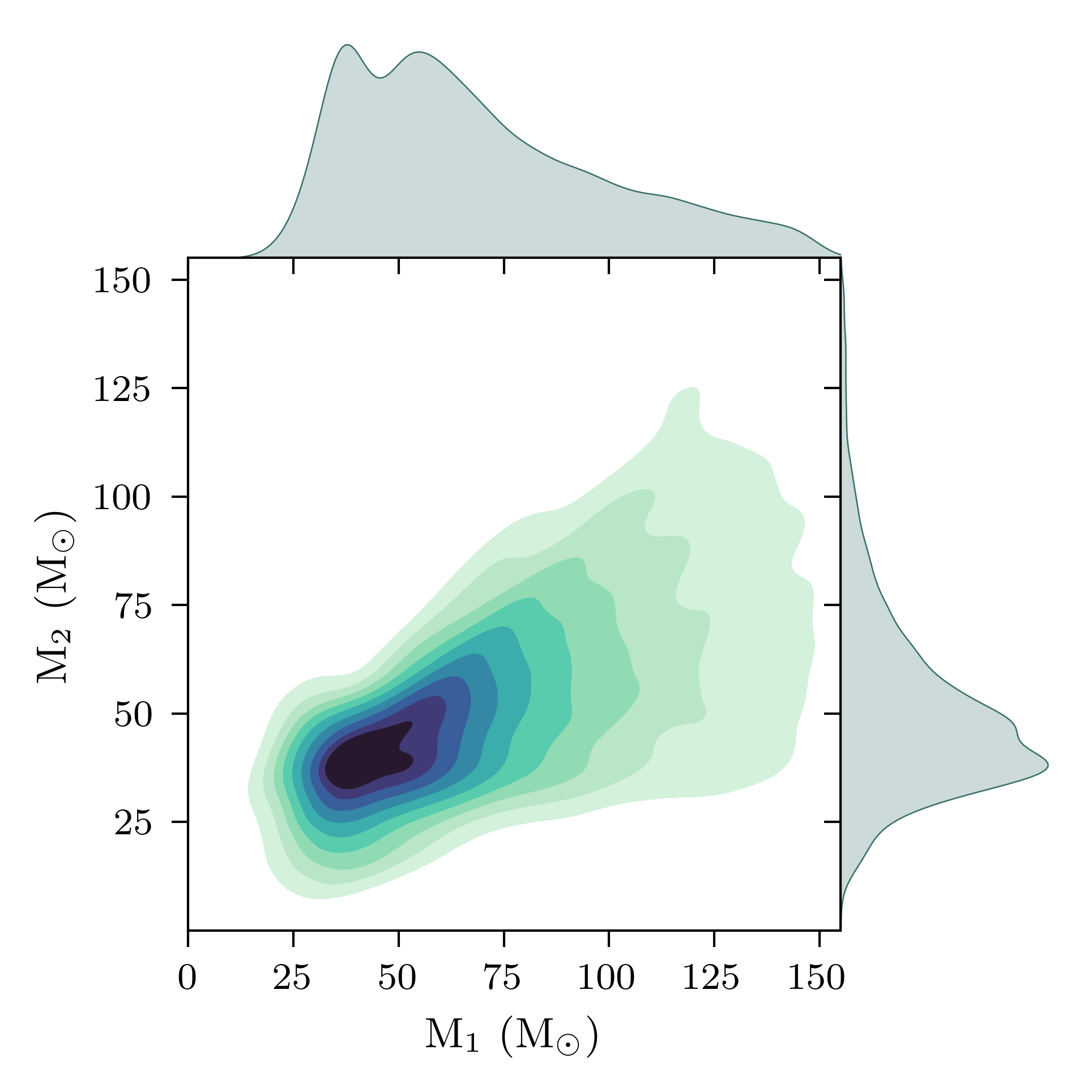}
 \caption{Distribution of masses ($ \mathrm{M}_2(\mathrm{M}_\odot) $ and $ \mathrm{M}_2(\mathrm{M}_\odot) $) of the eccentric initial binaries that later evolved into BBH or BH-NS. The concentration of the primary and secondary stars is defined by colour depth.}
\label{mass}
\end{figure}

To obtain a sufficient population of first Gen COs, we use a catalogue of N=10$^6$ initial binaries. Since our study focuses on BBH and BH-NS binaries, we restrict our analysis to initial binaries where at least one star evolves into a BH or NS.
As seen in Figure \ref{mass}, the initial mass exhibits a concentration near 44 M$\odot$ for the primary component and while for the secondary, it is approximately 36 M$\odot$.

\begin{figure}[htb!]
  \centering
	\includegraphics[width=8cm]{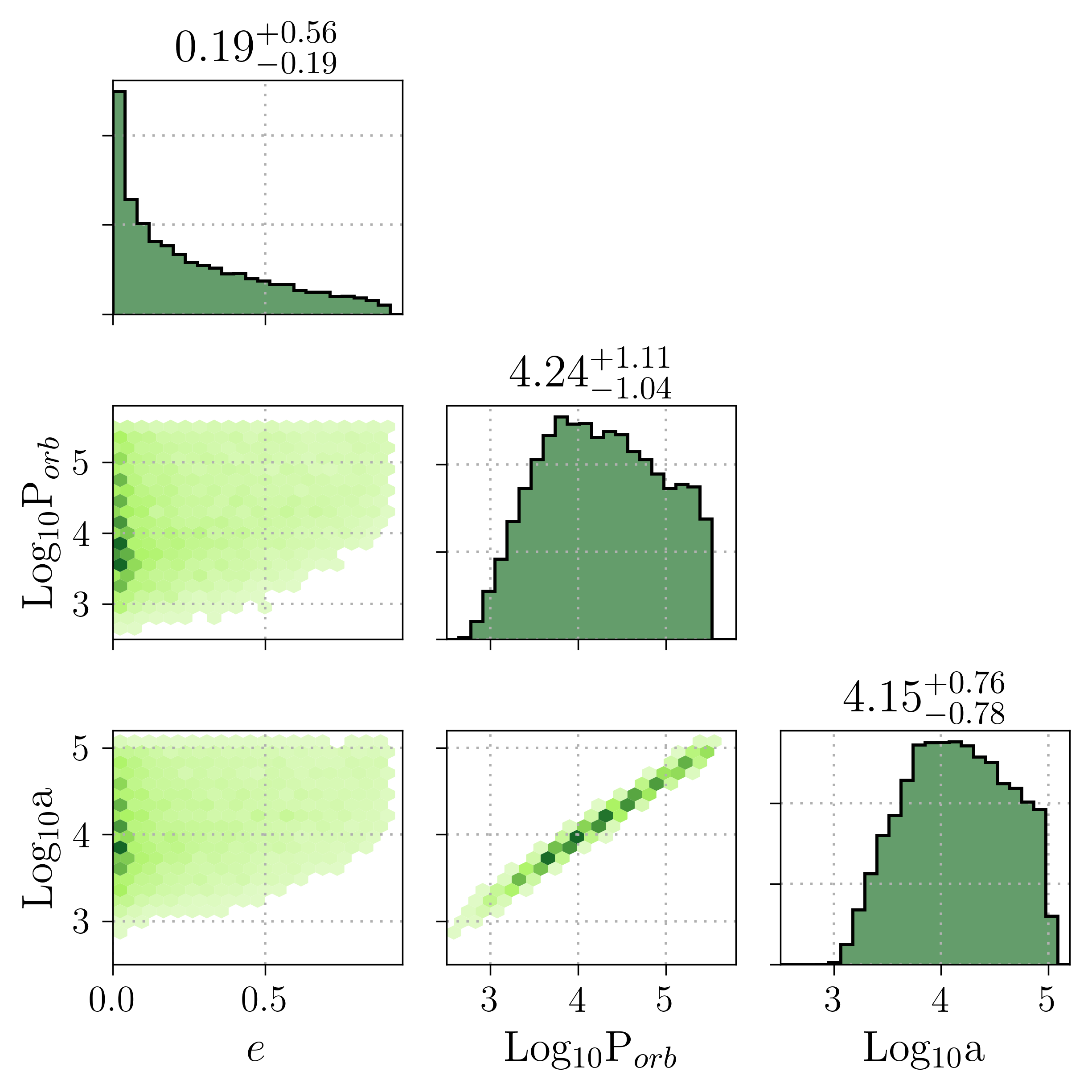}
	\caption{Distribution of the eccentricity, orbital period and separation of the progenitors resulting in CO. Along the diagonal, there are one-dimensional marginalised distributions for each parameter. Two-dimensional maps present the distributions over alternative parameters. Colour depth defines the concentration.}\label{corner}
\end{figure}

We use the Salpeter IMF, which favours a larger population of lower-mass stars. However, more massive stars have a higher likelihood of collapsing into BHs, resulting in a merging population with metallicities skewed toward lower values. Most stellar objects are concentrated in the low-metallicity regime ($\sim$0.002–0.004), though some progenitor binaries are effectively metal-rich ($\sim$0.007). The metallicity range in our study is broadly consistent with that of the Small Magellanic Cloud (SMC, $\mathrm{Z}_{\mathrm{SMC}} \gtrsim 0.0021$) and the Large Magellanic Cloud (LMC, $\mathrm{Z}_{\mathrm{LMC}} \gtrsim 0.0047$) \citep{met_data1,met_data}. From the simulation outcomes, it can be observed that the initial binaries have more population towards eccentricity $ e\approx 0.10$. The corner plot (Figure \ref{corner}) illustrates the distribution of selected intrinsic parameters of the initial binaries of our simulation. Our simulation shows that the maximum orbital period and separation are around 19.5 years and 350 AU, respectively, of the progenitors.

\subsection{First generation COs and corresponding strain}

\begin{figure}[htb!]
  \centering
	\includegraphics{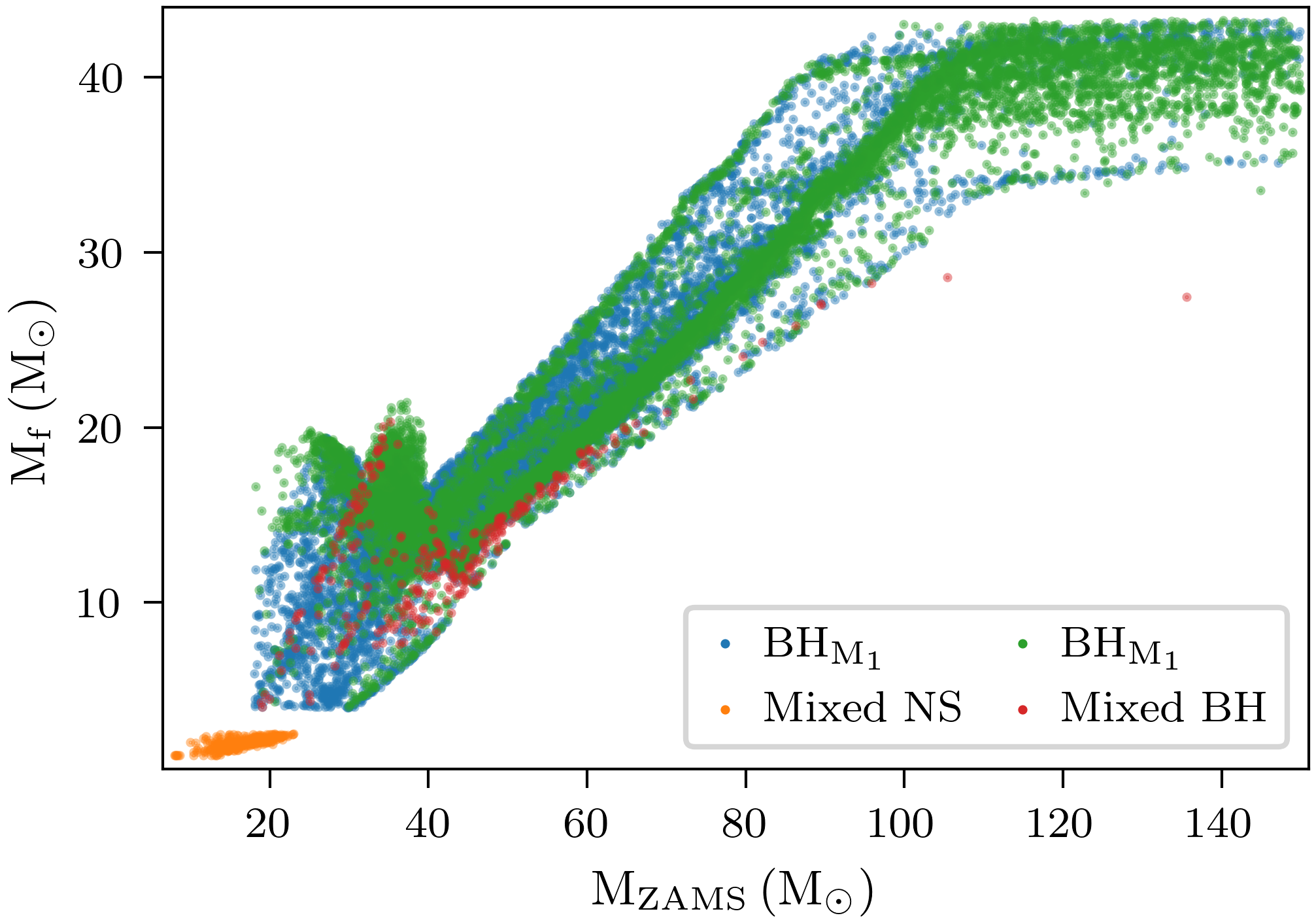}
	\caption{Final mass of CO ($\mathrm{M}_f$) vs initial stellar mass ($\mathrm{M}_{ZAMS}$) from our BPS simulation. The mass of the primary BH in a BBH system is denoted as BH$_{\mathrm{M}_1}$, while the secondary BH's mass is represented as BH$_{\mathrm{M}_2}$. In a BH-NS binary, the term `Mixed BH' refers to the BH component, whereas `Mixed NS' designates the NS within the system.}
	\label{mf_zsep}
\end{figure}

\begin{figure}[h!]
    \centering
    \includegraphics{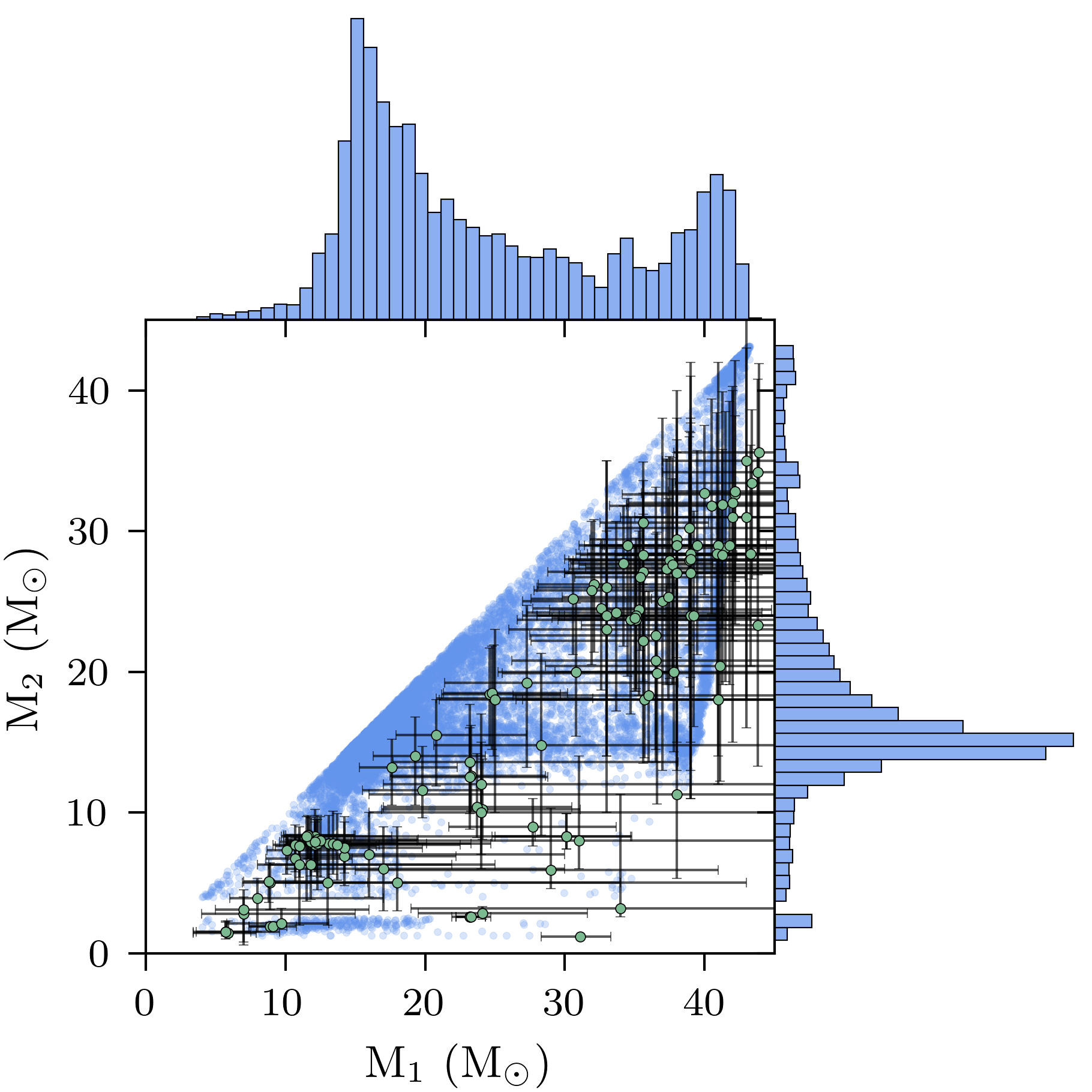}
    \caption{Distribution of primary and secondary masses of first Gen COs derived from simulations. The density of simulated CO masses is indicated by the scatter plot and histograms along each axis. Overlaid are observational data points from the LVK collaboration, showing primary and secondary masses of binary COs detected, which serve as a comparative benchmark for the simulation outcomes.}
    \label{final_mass}
\end{figure}

The core of a ZAMS star contracts after the nuclear burn is complete and is expected to evolve into a BH or a NS \citep{Bromm3}. The variation of ZAMS masses and the estimated gravitational mass of COs is shown in Figure \ref{mf_zsep}. We note that the number of BBH formed is much higher than the BH-NS binaries. In addition, BH masses in BH-NS binaries are significantly lower than in BBH binaries. In our simulation, the maximum mass of the BH reaches $\approx$ 43 M$_\odot$. The progenitor stars that evolve into NSs exhibit a maximum initial mass $\approx$ 23 M$_\odot$.\,\,The results indicate that in certain instances the secondary star becomes more massive than the primary star due to accretion. We noted that the primary and secondary masses of the COs predicted by our simulation are comparable to the masses of a subset of the detected GW events by the LVK collaborations (if the secondary mass exceeds the primary mass, the two are swapped to facilitate comparison with LVK observational results.), as shown in Figure \ref{final_mass}.

\begin{figure}[htb!]
    \centering
    \includegraphics{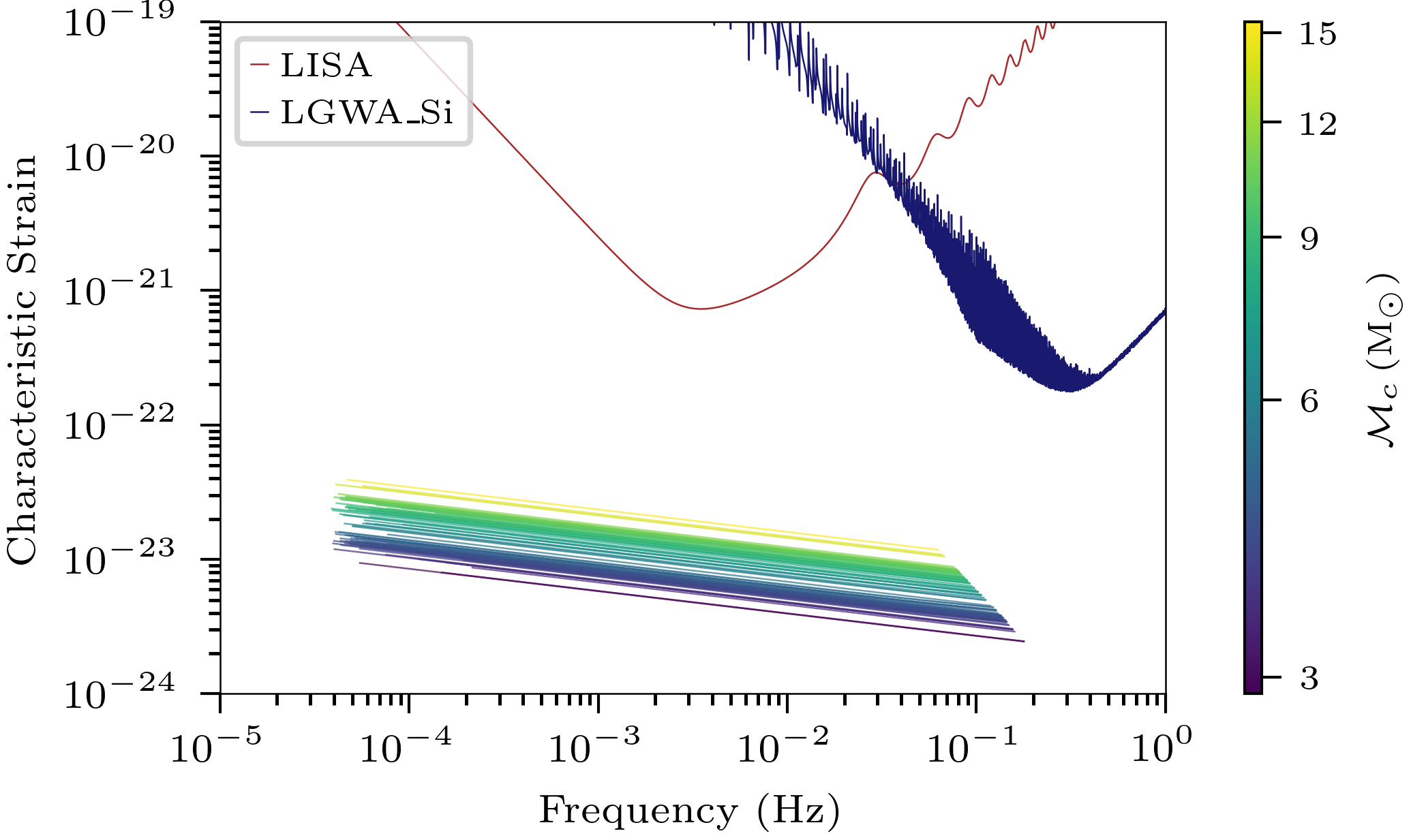}
    \caption{Characteristic strain against the frequency of COs formed in our simulation, at redshift $z$=4. We concentrated only on the inspiralling COs. The strain resulting from each CO is coloured by the chirp mass in the detector frame. The sensitivity curves of LISA and LGWA are represented by the left and right curves, respectively.}
    \label{gw}
\end{figure}

With the remnant mass distribution of the simulated CO binaries, the subsequent analysis focuses on characterising their GW emission properties. Numerous studies \citep{path1,path2} consider distinct pathways for the formation and evolution of COs, incorporating varying assumptions regarding a broad array of physical processes across different scales. These complexities contribute to significant challenges in predicting the mergers of first Gen COs \citep{Liu:2024mkh}. Despite the extensive versatility and wide range of detectability, in this study, we specifically focused on inspiralling binaries. Following the formalism in Eq. (\ref{e7}), we compute the characteristic strain for each inspiralling system, concentrating on the sub-Hz band relevant to future space-based detectors, LISA \citep{lisa}and LGWA \citep{lgwa} as shown in Figure \ref{gw}.

The signal-to-noise ratio (SNR) quantifies the strength of the strain. For eccentric sources, the total SNR is,

\begin{equation}
    \textrm{SNR}^2 =  \sum_{n=1}^ \infty \int_{f_{start}}^{f_{end}} \frac{h_{c,n}^2(f_n)}{f_n S_n(f_n)} d\ln{f_n}
\end{equation}

Where $S_n(f_n)$ is the power spectral density of the detector. We compared the resulting spectra from our model with the projected sensitivities of the advanced detectors. In this study, the observation period is considered to be four years. The sensitivity predicted by our model is of the order of $\mathcal{O}(10^{-2})$ for LISA and LGWA. 

\section{Conclusion and discussion}
\label{sec4}
In this study, we have modelled the formation and evolution of first-generation CO binaries originating from ZAMS stars in low-metallicity environments representative of the early Universe. By moving beyond the common single-metallicity assumption and instead assigning metallicities to each ZAMS star individually—using a mass–redshift–metallicity relation—we implemented a more realistic treatment within the COSMIC binary population synthesis framework. The final mass distribution of compact stellar remnants from the simulated catalogue of BBHs and BH–NS binaries with individual metallicities exhibits much higher concordance with observational data. Observations of binaries containing a large BH accreting from a Wolf-Rayet star and BPS show that the formation rate of binaries containing BHs significantly increases with decreasing metallicity \citep{di_co,wr1,wr2}.

Binary CO systems comprising double BHs are most commonly found in environments with low metallicity \citep{bh1,bh2}.
The larger number of BH and NS in our data can, therefore, be attributed to smaller values of metallicity set by the proposed M-Z relation.
In COSMIC, it is possible to assign an even lower constant metallicity to the entire population. This could potentially lead to an increase in the number of BH and NS compared to our current simulations, which follow the M-Z relation. However, metallicity is crucial in determining the evolutionary path of stars and binaries. To properly analyse even lower metallicity regimes, it would be necessary to consider more massive stellar progenitors, so as to remain consistent with the established correlation between stellar mass and metallicity. It is currently beyond the scope of our work. Analysis of the GW characteristic strain shows that a substantial fraction of the resulting inspiralling CO binaries emit chiefly in the sub-Hz regime. Yet the predicted strains typically lie beneath the anticipated sensitivities of forthcoming detectors such as LISA and LGWA.

Catalogues generated from the popsynth code, which provide comprehensive stellar populations rather than simple distributions, can be valuable resources for studying the stochastic gravitational wave background of clusters. With the advancement of observational data, BPS codes will become essential for conducting precise and detailed analyses of the stochastic gravitational wave background. Currently, the code does not account for evolving metallicity throughout stellar evolution. Introducing metallicity evolution, along with the progression from ZAMS stars to COs, is essential for improving the accuracy of population synthesis models in simulating realistic stellar evolution outcomes and binary interactions. With the advent of next-generation observatories and the consequent expansion of GW event samples- particularly at higher redshifts and in metal-poor environments-such advanced modelling will be essential for bridging theoretical frameworks, numerical simulations, and observational data in the era of multi-messenger astrophysics.

\begin{acknowledgments}
SRC extends gratitude to Maxim Khlopov of APC for valuable discussions during the early stages of this study. Authors also express their sincere thanks to Ranjini Mondol of IISc for insightful discussions and comments that have enhanced the quality of the manuscript. Authors are also thankful to Saibal Ray and Sourav Mitra for their valuable suggestions on improving the structure of the manuscript.
\end{acknowledgments}

We gratefully acknowledge the use of NumPy \citep{numpy}, SciPy \citep{scipy}, Matplotlib \citep{mat}, Seaborn \citep{seaborn}, Corner \citep{corner}, and COSMIC \citep{cosmic_software}.

\begin{appendix}
\section{Results with an alternative approach}\label{appen}
In addition to assigning individual metallicities to each stellar component within a binary system, as described in the main text, we also consider an alternative approach in which both components are assumed to share the same metallicity.
This scenario is motivated by the possibility that some binaries formed from the same chemically homogeneous environment in a stellar cluster.
To determine a representative stellar mass for computing the shared metallicity, we adopt the reduced mass formalism:
\begin{equation}
    \tilde{\mathrm{M}}= {\mathrm{M}_1 \mathrm{M}_2}/{(\mathrm{M}_1 + \mathrm{M}_2)}\label{eq:reduced_mass}
\end{equation}

where $\mathrm{M}_1$ and $\mathrm{M}_2$ are the masses of the primary and secondary stellar components, respectively. The reduced mass provides a physically meaningful quantity commonly employed in two-body dynamics to describe the relative motion of interacting masses.
In this context, it serves as an effective mass scale that reflects the coupled influence of both stars and is well-suited for use in empirical mass–metallicity relations.

We modified the original sampling function to allow \verb|metallicity| to accept both \verb|string| and \verb|float| values.
Setting a constant float value, sets the given metallicity value for all systems in the cluster.
Setting \verb|metalllicty| to \verb|"individual"| or \verb|"reduced_mass"| triggers the program to use Eq. \ref{e1} for calculating the metallicity(ies).
In case of reduced mass, binaries a system share the metallicity, $Z(\tilde{M})$ where $\tilde{M}$ comes from the result of Eq. \ref{eq:reduced_mass}.

\begin{figure}[htb!]
  \centering
	\includegraphics[scale=.85]{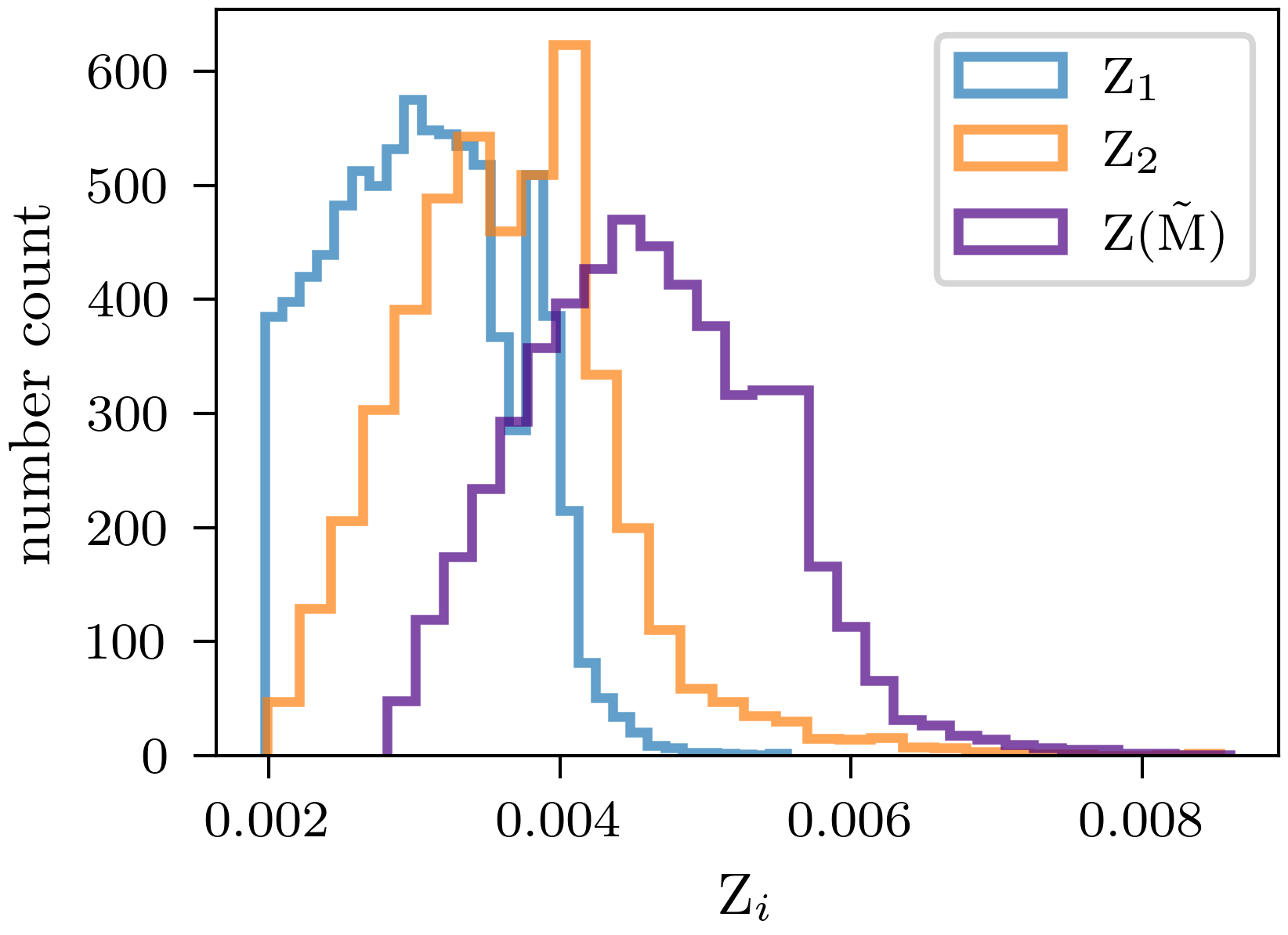}
	\caption{The distributions of the primary ($Z_1$) and secondary ($Z_2$) progenitors’ metallicities of the initial binaries, shown for individual metallicities and for the case of shared metallicities $Z(\tilde{M})$.
}
	\label{fig:example1}
\end{figure}

In our simulations, reduced mass systems exhibit consistent evolutionary trends concerning mass loss over time and the number of binaries that evolve into binary black holes. For systems formulated with a single metallicity per system, reduced mass metallicity can effectively reconcile the distinction between assigning individual metallicities and adopting a uniform metallicity across the stellar cluster.

\end{appendix}

\bibliography{meta_gw}

\end{document}